\documentclass[conference]{IEEEtran} 
\ifCLASSINFOpdf
\else
\fi
\usepackage{amsmath,epsfig}
\usepackage{amssymb}
\usepackage{bm}
\usepackage{algorithm}
\usepackage{algorithmic}
\usepackage{psfrag}
\usepackage{epsfig}

\usepackage[left=0.75in,right=0.7in,top=1in]{geometry}

\DeclareMathOperator*{\argmin}{argmin}
\DeclareMathOperator*{\argmax}{argmax}

\newcommand{\B}[1]{\boldsymbol{#1}}

\title{MIMO Beampattern and Waveform Design with Low Resolution DACs}
\author{\IEEEauthorblockN{Amine Mezghani and Robert W. Heath, Jr.}
\IEEEauthorblockA{Wireless Networking and Communications Group\\
Department of ECE, The University of Texas at Austin\\
Austin, TX 78712, USA\\ 
Email: \{amine.mezghani, rheath\}@utexas.edu} 
} 

\begin{document}
%
\maketitle
\begin{abstract} 
Digital beamforming and waveform generation techniques in MIMO radar offer enormous advantages in terms of flexibility and performance compared to conventional radar systems based on analog implementations. 
To allow for such fully digital design with an efficient hardware complexity, we consider the use of low resolution digital-to-analog converters (DACs) while maintaining a separate radio-frequency chain per antenna.  A  sum of squared residuals (SSR) formulation for the beampattern and spectral shaping problem is solved based on the Generalized Approximate Message Passing (GAMP) algorithm.
 Numerical results demonstrate good performance in terms of spectral shaping as well as cross-correlation properties of the different probing waveforms even with just 2-bit resolution per antenna.\let\thefootnote\relax\footnotetext{This research was supported by the U.S. Department of Transportation through the Data-Supported Transportation Operations and Planning (DSTOP) Tier 1 University Transportation Center and by the Texas Department of Transportation under Project 0-6877 entitled ``Communications and Radar- Supported Transportation Operations and Planning (CAR-STOP)" and a gift by Qualcomm. This material is also based upon work supported in part by the National Science Foundation under Grant No. CNS-1731658.}
\end{abstract}
%
%
\section{Introduction}
Achieving higher spatial resolution and offering higher degree flexibility for interference mitigation are main objective of future automotive radar  with multiple-input multiple-output (MIMO) antenna systems \cite{Fishler_2004}. Due to the next-generation vehicle requirements of Gbps communication data rate, it also desired  to reuse the same radar radio-frequency (RF) frontend jointly for wireless connectivity, particularly at millimeter-wave (mmWave) frequencies \cite{KumChoPre:IEEE-802.11ad-based-Radar::17,KumEltHea:Sparsity-Aware-Adaptive-Beamforming:18, GroLopVen:Opportunistic-automotive-radar:17}. State-of the-art radar systems, though, still rely on analog preprocessing such as the waveform generation and the beam steering, to alleviate the speed requirements on the analog-to-digital (ADC) and digital-to-analog converters (DACs). That approach does not allow for joint communication and sensing, not to mention offers little flexibility in dealing with interference issues and resource management. 

Fully digital radars offer the highest levels of flexibility. Unfortunately, with wide bandwidths and high resolution, DAC and ADCs become a power consumption bottleneck. To enable a practical low cost/power implementation with large bandwidth, we proposed in \cite{Kumari_2018} the use of low resolution ADCs, even down to one-bit per in-phase and quadrature components, while heavily relaxing the requirements (preamplifier gains, linearity, etc.) on the entire RF frontend. Our prior work though did not consider the companion problem of how to design the transmit waveform with a low resolution DAC. Thereby,  The waveform generation and beampattern matching problems are in fact critical tasks under these hardware constraints. These tasks aims at sending probing waveforms toward certain directions of interest with low autocorrelation sidelobes and  minimal cross-correlations values.


 Prior work has mainly focused on the MIMO beampattern design problem with constant envelope constraint \cite{Aldayel_2017,Wang_2012,Guo_2015}. There is  also some limited work on spectrally shaped binary sequence design, but only for a single dimension \cite{Mo_2015}. To the best of our knowledge, no prior work has tackled the spectral shaping and beamforming problems jointly with DAC constraints. In the context of MIMO communications, precoding with low resolution DACs has also gained attention as a mean to reduce hardware complexity \cite{Mezghani_2008_G,Jacobsson_2016,Jedda_2017}.  

In this paper, a spatial and temporal waveform co-design criterion based on the squared deviation to a desired reference unconstrained design is formulated for wideband MIMO with low resolution DACs and constant envelope signals.  We provide an efficient method for (suboptimally) solving the resulting non-convex optimization using the min-sum Generalized Approximate Message Passing (GAMP) algorithm. Numerical evaluations show that the resulting spectral shape and beampattern can be close the the ideal unconstrained reference  with nearly flat spectrum at the passband and small magnitudes in the stopband. Additionally, the constructed waveforms turn to have low cross-correlations values toward different directions.

\emph{Notation:}   The operators $(\bullet)^\mathrm {T}$, $(\bullet)^\mathrm {H}$, $\textrm{tr}(\bullet)$ and $(\bullet)^*$ stand for transpose, Hermitian (conjugate transpose), trace, and complex conjugate.  The notation $\B{I}_N$ represent the identity matrix of size $N$.  We represent the Hadamard (element-wise) and the Kronecker product of vectors and matrices by the operators "$\circ$" and "$\otimes$", respectively. Additionally, the operator ${\bf Diag}(\cdot)$ constructs a diagonal or a block diagonal matrix from a vector or a sequence of matrices. Further,  $\B{F}_{T}$ represents the normalized DFT matrix of size $T$ with swapped left and right halves, i.e., the zero-frequency component is at the center of the matrix. We define the projection operator on a certain set $\mathcal{S}$ as
\begin{equation}
{\rm prox}_{\mathcal{S}}(v)= \argmin_{x \in \mathcal{S}} |x-v|^2.
\end{equation}
Partial derivatives of functions with multiple complex arguments are expressed in terms of Wirtinger derivatives with respect to the first argument as  
\begin{equation}
f'(v,\cdots)=\frac{1}{2} \left(\frac{\partial f(v,\cdots)}{\partial {\rm Re}(v)} -{\rm j} \frac{\partial f(v,\cdots)}{\partial {\rm Im}(v)}   \right).
\end{equation}. 

%
%



\section{System model and problem formulation}

\begin{figure}[h]
\centerline{\includegraphics[width=8cm]{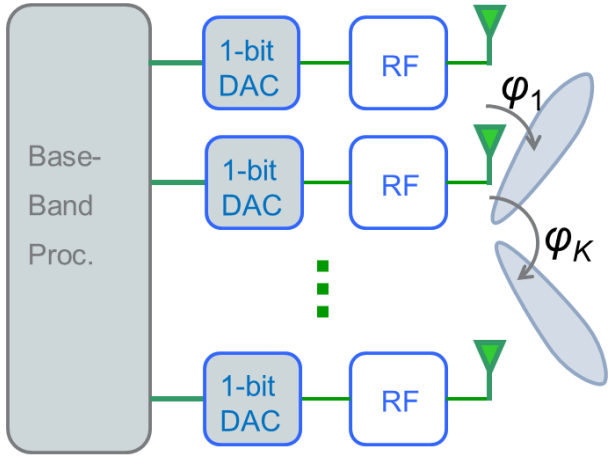}}
\caption{Digital beamforming architecture at the transmitter.}
\label{system_model}
\end{figure}

Consider a radar system with fully digital beamforming. It aims at simultaneously estimating ranges of certain surrounding targets at given directions, $\varphi_1,\cdots,\varphi_K$ as shown in Fig.~\ref{system_model}. Adopting the canonical model  from the literature, we assume the TX antenna array is a uniform linear array (ULA) of $N$ hypothetical isotropic antennas with half-wavelength spacing at the center frequency $f_{\rm c}$. The transmitted signal vector sequence in the discrete time domain is denoted by $\tilde{\B{x}}[0],\ldots,\tilde{\B{x}}^{\rm T}[T-1]$ with $T$ being the number of samples. Due to the constant envelope and  $b$-bit phase-only DAC constraint, the stacked baseband vector $\B{x}=[\tilde{\B{x}}^{\rm T}[0],\ldots,\tilde{\B{x}}^{\rm T}[T-1]]^{\rm T}$ is restricted to the set  
\begin{equation}
\begin{aligned}
\B{x} \in   \mathcal{X}^{TN},
\end{aligned}
\end{equation}
with the set of phase-only DAC values   
 \begin{equation}
  \mathcal{X}=\left\{{\rm e}^{{\rm j} \frac{2\pi}{2^b} (\ell +\frac{1}{2})}:  \ell=0,\cdots,2^{b-1}  \right\}.
    \label{set_X}
 \end{equation}
  The  broadband array frequency response in the $T$-DFT domain assuming all the frequencies propagate with the same speed is \cite{Brady_2015}
\begin{equation}
\begin{aligned}
&\B{a}(m,\varphi)\!=\!\!\left[1~ {\rm e}^{-{\rm j}{\pi} \cos(\varphi) (\frac{m}{TT_sf_{\rm c}}+1  )}\cdots  \right.  \\
& \quad\quad\quad\quad\quad\quad\quad\quad  \left. {\rm e}^{-{\rm j}{\pi} \cos(\varphi) (N-1) (\frac{m}{TT_sf_{\rm c}}+1  ) } \right]^{\rm T},
\end{aligned}
\label{array_res}
\end{equation}
where $f_{\rm c}$ is the center frequency, $T_s$ is the sampling interval and $m = -\frac{T}{2},\cdots,\frac{T}{2}-1  $ is the normalized frequency index .

Consider $K$ probing directions $\varphi_1,\cdots,\varphi_K$ for the desired beampattern as shown in Fig.~\ref{system_model}. We evaluate the following array response at the target directions to form the following matrix at each frequency index
\begin{equation}
\begin{aligned}
\B{A}[m]=\left[\B{a}(m,\varphi_1) \cdots \B{a}(m,\varphi_K) \right]^{\rm T}.
\end{aligned}
\end{equation}
Hence, the radiated field at the directions of interest can written in the DFT domain as
\begin{equation}
  \B{y}=\left[\cdots \tilde{\B{y}}[m]^{\rm T}  \cdots \right]^{\rm T} =\underbrace{{\bf Diag}(\cdots ,\B{A}[m], \cdots ) (\B{F}_{\!T} \otimes \B{I}_{\!N})}_{\B{B}} \B{x}.
\end{equation}
Given a certain desired beampattern intensity vector $\B{d}[m] \in \mathbb{R}_+^{K}$ for the $K$ probing  directions  at each frequency point $m$, the joint beampattern and waveform matching problem can be formulated as
\begin{equation}
\begin{aligned}
&\argmin\limits_{\B{s},\B{x},\beta} \left\| \beta \B{B} \B{x}  -    \B{s} \right\|_2^2 + \alpha \beta^2  ~ {\rm s.t.} ~ |s_j|^2 =d_j,~ \forall j,~ \B{x} \in \mathcal{X}^{NT}\!,
\end{aligned} 
\label{MSE_form}
\end{equation} 
 where  $\beta$ is a scaling factor accounting for the normalization of the stacked beampattern vector $\B{d}=\left[\cdots \B{d}[m]^{\rm T}  \cdots \right]^{\rm T}$. The regularization term $\alpha \beta^2 $ with parameter $\alpha$ aims at keeping $\beta$ minimal. In other words, this regularization controls the radiated power and the energy efficiency. Typically, waveforms with impulse-like autocorrelation are desired, therefore, $\B{d}[m]$  shall be flat within the allocated band (e.g. for $-\frac{T}{4}\leq m < \frac{T}{4}$) and zero elsewhere. For fixed $\B{x}$, we can solve the optimization (\ref{MSE_form}) with respect to $\B{s}$ and $\beta$ in closed form, i.e.,
 \begin{equation}
\begin{aligned}
 \B{s}=\sqrt{\B{d}} \circ  {\rm e}^{{\rm j} \cdot {\rm arg}(\B{B} \B{x})},
 \end{aligned} 
 \label{beta1}
\end{equation} 
and
\begin{equation}
\begin{aligned}
\beta  = \frac{\B{s}^{\rm H} \B{B} \B{x}}{ \|  \B{B}\B{x} \|_2^2+\alpha }.
\end{aligned}
\label{beta2} 
\end{equation}  
  The optimization with respect to $\B{x}$ is however NP-hard. In the following section, we provide an efficient algorithm for sub-optimally solving the non-convex problem (\ref{MSE_form}).     
\section{Proposed algorithm}
To solve the radar beampattern matching algorithm problem, we propose to use the iterative Generalized Approximate Message Passing (GAMP) algorithm \cite{rangan}. The min-sum-GAMP algorithm is intended to solve optimization problems involving a sum of numerous similar terms and a linear mixing of the vector to be optimized, as in (\ref{MSE_form}). Due to its astonishing performance that is reported in the literature, we adopt it for sub-optimally solving this non-convex problem.
     
The GAMP algorithm was derived in slightly different forms in several previous works \cite{rangan,Mezghani_WSA_2012,parker1} and it is provided in the appendix for convenience.  For more historical insights, we also encourage interested readers to consider the much earlier original works in statistical physics \cite{Mezard_1989}, where the GAMP iteration is referred to as the  Thouless-Anderson-Palmer (TAP) equations. The pseudo-code of GAMP is provide in Algorithm~\ref{GAMP_alg}. The recursive approach breaks apart the entire optimization problem into smaller scalar optimizations described by the function  
\begin{equation} 
   f_\ell(v,\xi^\ell)=  \argmax \limits_{x \in \mathcal{X}}   -\frac{1}{\xi^\ell} \| \xi^\ell x-v \| ^2 = {\rm prox}_{\mathcal{X}} \left( \frac{v}{\xi^\ell} \right)
   \label{f_ell}
\end{equation} 
that incorporate the constraint on $\B{x}$, and  the function 
\begin{equation} 
   g_\ell(-u,d,\theta^\ell)= \frac{1}{\theta^\ell} \argmax\limits_w  \left[ \rho(w,d) -\frac{1}{\theta^\ell} \| w-u\|^2 \right] - \frac{u}{\theta^\ell}
 \label{out_func}
\end{equation} 
that is related to the cost function. More precisely, we have
\begin{equation} 
 \rho(w,d)  =  - \min\limits_{|s|^2 = d}  \|  w-s \| ^2=  - \| w-\sqrt{d}  {\rm e}^{{\rm j} \cdot {\rm arg}(w)}  \| ^2, 
\label{rho_w}
\end{equation} 
according to the formulation (\ref{MSE_form}). Solving the minimization in (\ref{out_func}), we get the closed form expression 
\begin{equation} 
g_\ell(-u,d,\theta^\ell) =  \frac{ \sqrt{d} {\rm e}^{{\rm j} \cdot {\rm arg}(u)} -u}{1+\theta^\ell}.
\label{g_ell}
\end{equation} 
The scalar functions $f_\ell$ and $g_\ell$ are applied element-wise to vectors in the GAMP algorithm, resulting into the so-called input and output steps, respectively.
The messages exchanged between the input and output steps consist, however, not only of the results of the individual scalar optimizations but also the curvature around these optima, which is crucial for faster convergence. The curvature message vectors $\boldsymbol{\xi}^{\ell}$ and $\boldsymbol{\theta}^{\ell}$  are obtained by means of the derivatives $f'_\ell$ and $g'_\ell$  with respect to the first argument. The Wirtinger derivative of the function $g'_\ell$ in (\ref{g_ell}) with respect to the first argument $-u$ reads as
\begin{equation} 
g'_\ell(-u,d,\theta^\ell) =  \frac{1}{1+\theta^\ell} -\frac{\sqrt{d}}{2|u|(1+\theta^\ell)}.
\end{equation}
The derivative  $f'_\ell$  of (\ref{f_ell}) vanishes almost everywhere due to the discreteness of the set $\mathcal{X}$, i.e.,  $\boldsymbol{\theta}^{\ell}=\B{0}$, which might affect the convergence behavior. 
To cope with this issue, we use the approximation
\begin{equation} 
f'_\ell(v,\xi^\ell) \approx  \frac{1}{2|v|},
\end{equation}
which becomes exact for the infinite resolution case with only a constant envelope constraint. The update for the scaling factor $\beta$  is done at each iteration based on (\ref{beta2}) and (\ref{beta1}). 
It should be noted that  the GAMP does not provide guarantees for optimality or convergence.  A damping strategy is used to enforce the convergence in our case as proposed in \cite{Rangan_GAMP_2014}.  

\begin{algorithm}[t]
\caption{Damped min-sum-GAMP algorithm}
\label{GAMP_alg}
\begin{algorithmic}[1]
\STATE \textbf{Input:}  desired pattern vector $\B{d}$, $\B{B}= {\bf Diag}(\cdots ,\B{A}[m], \cdots ) (\B{F}_{\!T} \otimes \B{I}_{\!N})$, $\alpha=K\cdot T$ 
\STATE \textbf{Initialize:} $\boldsymbol{z}^0=\boldsymbol{0}$,
$\boldsymbol{x}^0=\boldsymbol{0}$, $\boldsymbol{\theta}^0={\rm
const}$, $\beta^{0}=1$
\\     
$ 0 \leq \mu < 1 $,  $\ell \leftarrow 0$      
\REPEAT
\STATE $\B{A} \! \leftarrow \! \beta^\ell \B{B}$
\STATE $\ell \! \leftarrow \!  \ell+1$
\STATE Output Step:\\
 $\boldsymbol{z}^{\ell} \! \leftarrow \! g_\ell(-\boldsymbol{A}\hat{\boldsymbol{x}}^{\ell-1}  + \boldsymbol{\theta}^{\ell-1} \circ\boldsymbol{z}^{\ell-1},\boldsymbol{d},\boldsymbol{\theta}^{\ell-1})$\\
$\boldsymbol{\xi}^{\ell} \! \leftarrow \! (\boldsymbol{A} \circ \boldsymbol{A}^*)^{\rm T}
g'_\ell ( -\boldsymbol{A}\hat{\boldsymbol{x}}^{\ell-1}  + \boldsymbol{\theta}^{\ell-1} \circ\boldsymbol{z}^{\ell-1}, \boldsymbol{\theta}^{\ell-1}, 
\boldsymbol{d}) $ \\
\STATE Input Step with damping: \\
$\boldsymbol{x}^{\ell}  \! \leftarrow \! (1-\mu)\boldsymbol{x}^{\ell}+ \mu f_\ell(\boldsymbol{A}^{\rm H}
\boldsymbol{z}^\ell  + \boldsymbol{\xi}^\ell \circ
\hat{\boldsymbol{x}}^{\ell-1},  \boldsymbol{\xi}^\ell)$ \\
$\boldsymbol{\theta}^\ell \! \leftarrow \! \mu  \cdot  (\boldsymbol{A} \circ \boldsymbol{A}^*) 
f'_\ell(\boldsymbol{A}^{\rm H} \boldsymbol{z}^\ell  +
\boldsymbol{\xi}^\ell \circ \hat{\boldsymbol{x}}^{\ell-1},  \boldsymbol{\xi}^\ell )$
\STATE Optimizing the scaling factor: \\
$\beta^{\ell}  \! \leftarrow \! \frac{(\sqrt{\B{d}} \circ  {\rm e}^{{\rm j} \cdot {\rm arg}(\B{B} \B{x}^{\ell})} )^{\rm H} \B{B} \B{x}^{\ell}}{ \|  \B{B}\B{x}^{\ell} \|_2^2+\alpha }$
\UNTIL{the cost does not significantly decrease or a maximum iteration count has
 been reached}
\STATE Final projection:\\
   $\B{x}_{\rm sol}={\rm prox}_{\mathcal{X}}(\B{x}^{\ell})$
\end{algorithmic}
\end{algorithm}
\section{Simulation Results}
\begin{figure}[t]
\centerline{\includegraphics[width=8cm]{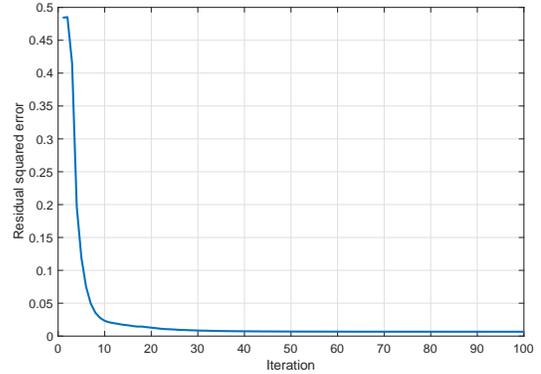}}
\caption{Convergence behavior of the GAMP algorithm with $N=128$ antennas, $T=1024$ samples and damping factor $\mu=0.3$.}
\label{sim_error}
\end{figure}
\begin{figure}[ht]
\centerline{\includegraphics[width=8cm]{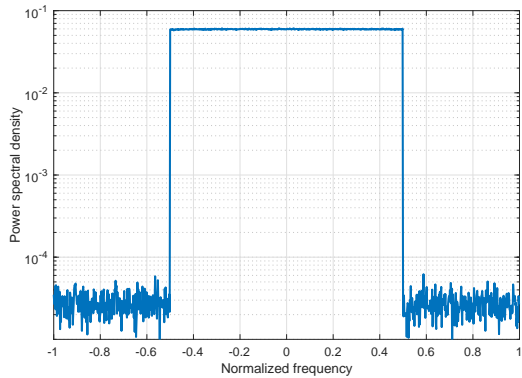}}
\caption{Power spectral density as function of frequency/1~GHz.}
\label{sim_psd}
\end{figure}
\begin{figure}[ht]
\centerline{\includegraphics[width=8cm]{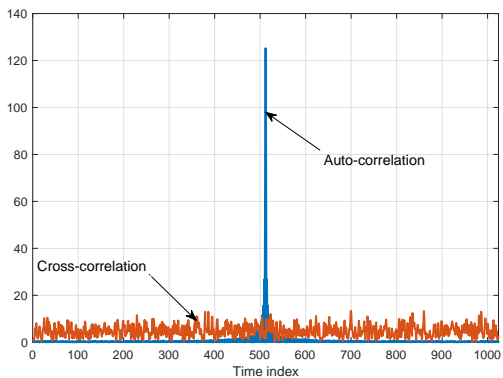}}
\caption{Auto- and cross-correlation values.}
\label{sim_corr}
\end{figure}
\begin{figure}[ht]
\centerline{\includegraphics[width=8cm]{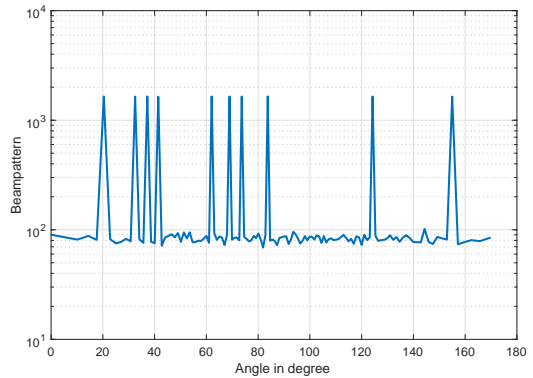}}
\caption{Radiation pattern for 10 target directions.}
\label{sim_pattern}
\end{figure}
The performance of the proposed joint beampattern and waveform design algorithm is investigated by means of simulations for the case of 2-bit DACs (i.e., 1-bit for each inphase and quadrature component). We consider an array of $N=128$ antennas, while the temporal sequence has $T=1024$ samples. Additionally, we assume a carrier frequency  of $f_c=77$ GHz, a bandwidth  of 1 GHz, and a sampling frequency of $1/T_s=2$ GHz. The desired beampattern points to $K=10$ predefined scanning  directions chosen between 0 and $\pi$ with equal intensity. The convergence behavior of the proposed GAMP is depicted in Fig.~\ref{sim_error}. The algorithm converges typically within 20 to 30 iterations. Fig.~\ref{sim_psd} shows the power spectral density which is very close to the desired rectangular shape.  In fact, the out-of-band (OOB) radiation level is more than 30dB  below the signal level. This is quite surprising given the extreme low resolution of the DACs. The resulting autocorrelation property is  therefore almost impulse-like as shown in Fig.~\ref{sim_corr}. In addition, Fig.~\ref{sim_pattern} shows that waveforms targeted at different  directions $\varphi_k$
\begin{equation}
  \B{y}_k = {\bf Diag}\left(\B{a}(-T/2,\varphi_k), \cdots, \B{a}(T/2-1,\varphi_k)\right) (\B{F}_{\!T} \otimes \B{I}_{\!N}) \B{x},
\end{equation}
$1 \leq k  \leq K$, exhibit low cross-correlation properties, which is very desired as well. Interestingly, this desired cross-correlation effect is obtained although it is not incorporated in the objective function  (\ref{MSE_form}). Finally, Fig.~\ref{sim_pattern} shows the antenna radiation pattern with pointy beams in the directions of interest.  It is worth mentioning that the sidelobe radiation could be actually enhanced by extending the objective function  (\ref{MSE_form}) to include the desired sidelobe behavior. The improvement will be however at the cost of higher OOB radiation. This trade-off could be investigated in future work.

\section{Conclusion}
An GAMP based algorithm is proposed to synthesize spatio-temporal waveforms for a MIMO radar systems under low resolution DACs and constant modulus constraint. 
To this end, an objective function is formulated that aims at concentrating the power toward certain desired angles and within a given allocated band. 
It is also possible to extend the objective function by additional terms to achieve higher suppression at the undesired angles.   
Simulation results show that the constructed waveforms exhibit high stopband attenuation and flat passband spectrum even with just 2-bit resolution per antenna. Therefore, even a severe spectral interference constraint can be incorporated while still keeping nearly ideal autocorrelation properties. Further, the different probing waveforms reflected back from the target angles have minimal cross-correlation values. A potential extension of the work is considering   joint multiuser MIMO communication-radar precoding and signal design  with low resolution DACs.       

\appendix
   For the derivation of the presented algorithm, assuming fixed $\beta$, a message passing algorithm  on factor graphs based on the min-sum  rules is considered while ignoring cycles. This approach is fairly similar to the derivation in \cite{rangan}. It is made possible due the structure of the cost function (\ref{MSE_form}) which consists of a sum of similar terms that are in turn functions of linearly mixed optimization variables. Therefore, the global optimization can be solved approximately by considering simpler individual optimizations. Each iteration $\ell$ of this algorithm consists namely in exchanging  messages (max-marginals) $\lambda_{j,i}^\ell(x_i)$ and $\pi_{j,i}^\ell(x_i)$  from each desired signal node $j \in \{1,\cdots, TK\}$ to
each sample node $i \in \{1,\cdots, TN\} $ (output step), and vice versa (input step). Specifically, the output step  messages are given by
\begin{equation}
\lambda_{j,i}^\ell(x_i)= \max_{\B{x} \backslash x_i }  \rho(\B{a}_{j}^{\rm T}\B{x},d_j) + \sum_{i' \neq i} \pi_{j,i'}^\ell(x_{i'}),
\label{horiz} 
\end{equation}
where $\rho(w,d)$ is defined in (\ref{rho_w}) and $\B{a}_{j}^{\rm T}$ is the $j$-th row of the matrix $\beta \B{B}$ for fixed $\beta$. The input step messages are then
\begin{equation}
\pi_{j,i}^\ell(x_i)= \log \mathbb{I}(x_i \in \mathcal{X}) + \sum_{j' \neq j} \lambda_{j',i}^\ell(x_i).
\label{verti}
\end{equation}
 The messages are initialized at $t=0$ with $\lambda_{j,i}^\ell(x_i)=0$, and the final estimate can be calculated as the maximizer of
\begin{equation} 
 \pi_{i}^\ell(x_i) = \log \mathbb{I}(x_i \in \mathcal{X}) + \sum_j  \lambda_{j,i}^\ell(x_i).  
\end{equation}
We now provide an approximation for the min-sum algorithm based on quadratic approximations of the messages around their corresponding maximizers, assuming a large matrix $\B{A}$ and that its entries $a_{j,i}$ are of the same order. To this end, we introduce the following values 
\begin{equation}
\begin{aligned}
  \hat{x}_i^\ell &= \argmax_{x_i} \pi_{i}^\ell(x_i)  \\
  \hat{x}_{j,i}^\ell&= \argmax_{x_i} \pi_{j,i}^\ell(x_i)  \\
  \frac{1}{\mu_i^{\ell}}&= - \frac{\partial^2}{\partial x_i \partial x_i^*} \pi_i^\ell(\hat{x}_i^\ell) \\
  \frac{1}{\mu_{j,i}^\ell} &= -\frac{\partial^2}{\partial x_i \partial x_i^*} \pi_{j,i}^\ell(\hat{x}_{i,j}^\ell). 
\end{aligned}
\label{mumu}
\end{equation}
Then, we can use the following quadratic approximation for the input step messages around its maximizer (neglecting derivatives w.r.t. $x^2$)
\begin{equation}
\pi_{j,i}^\ell(x_i) \approx  \pi_{j,i}^\ell(\hat{x}_{j,i}^\ell ) - \frac{1}{ \mu_{i}^\ell } |x_i- \hat{x}_{j,i}^\ell |^2,
\end{equation}
where we assume $\mu_{j,i}^\ell \approx \mu_{i}^\ell$  in the large dimensional case. Thus,  the cost function in (\ref{horiz}) becomes 
\begin{equation}
   \rho(\B{a}_{j}^{\rm T}\B{x},d_j) + \sum_{i' \neq i} \pi_{j,i'}^\ell(\hat{x}_{j,i'}^\ell ) - \frac{1}{ \mu_{i'}^\ell } |x_{i'} - \hat{x}_{j,i'}^\ell |^2.
\end{equation}
The motivation for such approximation is that the impact of $x_i$ on the function $ \rho(\B{a}_{j}^{\rm T}\B{x},d_j)$  is asymptotically very small and that the maximizing value of $x_i$ for the cost function (\ref{horiz}) will  not deviate much from $\hat{x}_{j,i}$.
We solve the maximization in (\ref{horiz}) in  two steps
\begin{equation}
\max_{q_j} \max_{\B{x} \backslash x_i, {\rm ~s.t.~}\B{a}_j^{\rm T}\B{x}=q_j}  \rho(q_j,d_j)  - \sum_{i' \neq i} \frac{1}{ \mu_{i'}^\ell } |x_{i'} - \hat{x}_{j,i'}^\ell |^2.
\end{equation}
The solution for the inner maximization leads to 
\begin{equation}
\max_{q_j}  \rho(q_j,d_j)  - \frac{1}{ \theta_{j,i}^\ell} |q_j - \hat{u}_{j,i}- a_{j,i} x_i|^2,
\end{equation}
where, we introduced the variables 
\begin{equation}
q_j= a_{j,i} x_i + \sum_{i' \neq i} a_{j,i'} x_{i'},
\end{equation} 
\begin{equation}
\hat{u}_{j,i}^\ell = \sum_{i' \neq i} a_{j,i'}  \hat{x}_{j,i'}^\ell,
\end{equation}
and
\begin{equation}
\theta_{j,i}^\ell= \sum_{i' \neq i} |a_{j,i'}|^2 \mu_{i'}^\ell \approx   \sum_{i} |a_{j,i}|^2 \mu_{i}^\ell =   \theta_{j} ~ \forall i,
\label{thetaj}
\end{equation}
where the approximation holds for the large system limit. 
Furthermore,  by defining
\begin{equation}
\hat{u}_{j}^\ell= \sum_i  a_{j,i} \hat{x}_{j,i}^\ell,
\label{uj}
\end{equation}
we get
\begin{equation}
\max_{q_j}   \rho(q_j,d_j) - \frac{1}{ \theta_{j}^\ell} |q_j - \hat{u}_{i}^\ell - a_{j,i} \hat{x}_{j,i}^\ell - a_{j,i} x_i|^2.
\end{equation}
These  obtained output step messages can be then approximated as
\begin{equation}
\begin{aligned}
\lambda_{j,i}^\ell(x_i) &= \max_{q_j}  \rho(q_j,d_j) -\frac{1}{ \theta_{j}^\ell} |q_j - \hat{u}_j^\ell + a_{j,i} \hat{x}_{j,i}^\ell - a_{j,i} x_i|^2  \\
 &=  \max_{q_j}  \rho(q_j,d_j)  -\frac{1}{ \theta_{j}^\ell} |q_j - \hat{u}_j^\ell - a_{j,i} (  x_i-\hat{x}_{j,i}^\ell) |^2 \\
  &\approx  \max_{q_j}  \rho(q_j,d_j)  -\frac{1}{ \theta_{j}^\ell} |q_j - \hat{u}_j^\ell - a_{j,i} (  x_i-\hat{x}_{i}^\ell) |^2, \\
\end{aligned}
\end{equation}
where the approximation is obtained by neglecting the terms of order $|a_{j,i}|^2$.
Let us now define 
\begin{equation}
\begin{aligned}
\hat{q}_j^\ell = \argmax_{q_j}  \rho(q_j,d_j)  -\frac{1}{ \theta_{j}^\ell} |q_j - \hat{u}_j^\ell  |^2, \\
\end{aligned}
\end{equation}
and
\begin{equation}
\begin{aligned}
z_j^\ell= \frac{1}{\theta_{j}^\ell} (\hat{q}_j^\ell - \hat{u}_j^\ell  )  \dot{=}  g_\ell(- \hat{u}_j^\ell,d_j).
\label{z_jt}
\end{aligned}
\end{equation}
As done before for the input step message, we derive now a second order expansion of the output step message around  $\hat{x}_{i}^\ell$. Evaluating the first derivative 
\begin{equation}
\begin{aligned}
\left. \frac{ \partial \lambda_{j,i}^\ell(x_i) }{ \partial x_i^*}  \right|_{x_i=\hat{x}_{i}^\ell}= a_{j,i} z_j^\ell,
\end{aligned}
\end{equation}
and the second derivative
\begin{equation}
\begin{aligned}
\left. -\frac{ \partial^2 \lambda_{j,i}^\ell(x_i) }{ \partial x_i \partial x_i^*}  \right|_{x_i=\hat{x}_{i}^\ell}  \!\!\!\!\!\!\!\!= - a_{j,i}  \frac{\partial z_j^\ell }{ \partial x_i}= |a_{j,i}|^2   g'_\ell(- \hat{u}_j^\ell,d_j)  \dot{=}|a_{j,i}|^2 \frac{1}{ \mu_{j}^{s,\ell}},
\end{aligned}
\end{equation}
leads to 
\begin{equation}
\begin{aligned}
\lambda_{j,i}^\ell(x_i) \approx  {\rm const~} -\frac{1}{ \mu_{j}^{s,\ell}} | \mu_{j}^{s,\ell} z_j^\ell - a_{j,i} (  x_i-\hat{x}_{i}^\ell) |^2. \\
\end{aligned}
\end{equation}
The input step messages can be obtained now as (c.f. (\ref{verti}))
\begin{equation}
\pi_{j,i}^\ell(x_i)= \log \mathbb{I}(x_i \in \mathcal{X}) -\frac{1}{ \xi_{j,i}^{\ell}} | \hat{v}_{j,i}^\ell- \xi_{j,i}^{\ell} x_i|^2,
\end{equation}
where we introduced the definitions
\begin{equation}
\begin{aligned}
\hat{v}_{j,i}^\ell &=  \sum_{l \neq j} ( a_{l,i}  z_i^\ell + \frac{1}{\mu_{l}^{s,\ell}} |a_{l,i}|^2  \hat{x}_{i}^\ell)    \\
                      &= \xi_{j,i}^{\ell} \hat{x}_i^\ell + \xi_{j,i}^{\ell} \sum_{l \neq j}  a_{l,i} z_l^\ell  \\
                      &= \underbrace{\xi_{j,i}^{\ell} \hat{x}_i^\ell +\sum_{l}  a_{l,i} z_l^\ell}_{\hat{v}_{i}^\ell}-   a_{j,i} z_j^\ell.
\label{vit}
\end{aligned}
\end{equation}
Thereby, we used the substitution
\begin{equation}
\begin{aligned}
\xi_{j,i}^{\ell} &= \sum_{l \neq j} |a_{l,i}|^2 \frac{1}{\mu_{l}^{s,\ell}} \approx  \sum_{j} |a_{j,i}|^2 \frac{1}{\mu_{l}^{s,\ell}} \dot{=}  \xi_{i}^{\ell}   ~\forall j,
\label{xit}
\end{aligned}
\end{equation}
where we can neglect the dependency on $j$. 
This corresponds to the calculation of $\B{\xi}^\ell$ in  the algorithm.
We can proceed  to compute $\hat{x}_{j,i}^\ell$
\begin{equation}
\begin{aligned}
\hat{x}_{j,i}^\ell  &=        \argmax_{x_i} \log \mathbb{I}(x_i \in \mathcal{X}) - \frac{1}{ \xi_{i,j}^{\ell}} |\hat{v}_{i,j}^\ell-  \xi_{i}^{\ell}  x_i|^2 \\
               &=        \argmax_{x_i} \log \mathbb{I}(x_i \in \mathcal{X}) - \frac{1}{ \xi_{i,j}^{\ell}} |\hat{v}_{i}^\ell -  a_{j,i} z_j^\ell -  \xi_{i}^{\ell}  x_i|^2 \\
               & \approx  \hat{x}_i^{\ell} -  \Gamma_i^\ell  a_{j,i} z_j^\ell, \\   
               \label{x_temp}
\end{aligned}
\end{equation}
where we define 
\begin{equation}
\begin{aligned}
\hat{x}_i^{\ell} &=\argmax_{x_i} \pi_i^\ell(x_i) \\
& = \argmax_{x_i} \log \mathbb{I}(x_i \in \mathcal{X}) -\frac{1} { \xi_{i}^{\ell}} |\hat{v}_{i}^\ell-  \xi_{i}^{\ell}  x_i|^2  \\
&\dot{=} f_\ell(\hat{v}_{i}^\ell),
\label{ft2}
\end{aligned}
\end{equation}
and
\begin{equation}
 \Gamma_i^\ell =\frac{\partial \hat{x}_i^\ell}{ \partial \hat{v}_i} = f'_\ell(\hat{v}_{i}^\ell).
 \label{dit}
\end{equation}
Next, it can be shown, assuming that the first derivative  ${\pi_i^{\ell}}'(\hat{x}_i^{t+1})=0$,  that $\Gamma_i^\ell = \mu_{i}^\ell $ (c.f. (\ref{mumu})).
Therefore, (\ref{thetaj}) becomes, while neglecting the dependency on the index $i$ 
\begin{equation}
\theta_{j}^\ell=   \sum_{i} |a_{j,i}|^2   f'_\ell(\hat{v}_{i}^\ell).
\label{thetaj2}
\end{equation}
Finally, (\ref{uj}) becomes using  (\ref{x_temp}) and the fact that $ \mu_{i}^\ell = \Gamma_i^\ell  = f'_\ell(\hat{v}_{i}^\ell)$ as follows
\begin{equation}
\hat{u}_{j}^\ell= \sum_i  a_{j,i} \hat{x}_{i}^\ell - \theta_{j}  z_j^\ell.
\label{uj2}
\end{equation}
It can be observed that the steps presented in the algorithm corresponds to Eqs (\ref{uj2}), (\ref{z_jt}),  (\ref{xit}), (\ref{vit}), (\ref{ft2}) and (\ref{thetaj2}).



\bibliographystyle{IEEEbib}
\bibliography{ref}

\end{document}